\documentclass[useAMS,usenatbib]{mn2e}
\usepackage{graphicx}
\usepackage{epstopdf}
\usepackage{amssymb}
\usepackage{amsmath}
\usepackage{subfigure}
\usepackage{url}
\usepackage{hyperref}

\title[Classifiers for {\it Fermi} AGN]{Gamma-ray active galactic nuclei type determination through machine-learning algorithms}
\author[Hassan et al.]{T. Hassan$^{1}$\thanks{E-mail:
thassan@gae.ucm.es}, N. Mirabal$^{1,2}$\thanks{E-mail:
mirabal@gae.ucm.es}, J. L. Contreras$^{1}$ and I. Oya$^{3}$\\
$^{1}$Dpto. de F\'isica At\'omica,
Molecular y Nuclear, Universidad Complutense de
Madrid, Spain\\
$^{2}$Ram\'on y Cajal Fellow\\
$^{3}$Institut f\"ur Physik, Humboldt-Universit\"at zu Berlin, Newtonstr. 15, D 12489 Berlin, Germany\\
}

\begin{document}

\date{}

\pagerange{\pageref{firstpage}--\pageref{lastpage}} \pubyear{2012}

\maketitle

\label{firstpage}

\begin{abstract}
The {\it Fermi Gamma-ray Space Telescope (Fermi)} is producing the most detailed inventory of the gamma-ray  
sky to date. Despite tremendous achievements approximately 25\% of all 
\textit{Fermi} extragalactic sources 
in the \textit{Second Fermi LAT Catalogue} (\textit{2FGL}) are listed as active galactic nuclei (AGN) 
of uncertain type. Typically, 
these are suspected blazar candidates without a conclusive optical spectrum or lacking spectroscopic observations. 
Here, we explore the use of machine-learning algorithms -- 
Random Forests and Support Vector Machines --
 to predict specific AGN subclass based on observed
gamma-ray spectral properties. 
After training and testing on identified/associated AGN from the 2FGL
we find that 235 out of 269 AGN of uncertain type have properties 
compatible with gamma-ray BL Lacs and flat-spectrum radio quasars with 
accuracy rates of 85\%. Additionally, direct 
comparison of our results with class predictions made after 
following the infrared 
colour-colour space of \citet{massaro} shows 
that the agreement rate is over four-fifths for 54 overlapping sources, 
providing independent 
cross validation. These results 
can help tailor follow-up spectroscopic programmes and inform future 
pointed surveys with ground-based Cherenkov telescopes.

\end{abstract}

\begin{keywords}
gamma-rays: observations -- galaxies: active
\end{keywords}

\section{Introduction}

The {\it Fermi} 
mission has revolutionised our knowledge of 
high-energy gamma ray sources in the 100 MeV to 100 GeV energy range. 
Instrumentally, 
the Large Area Telescope (LAT)  
increased sensitivity and sky coverage represents 
a giant leap forward compared to  
EGRET \citep{EGRET}, allowing access to the gamma-ray 
sky with unprecedented detail. 
With over two years of collected data, 
the Second {\it Fermi} LAT Catalogue 
(2FGL) lists a total of 1873 point-like sources, including 1092 objects 
connected with known AGN at other wavelengths \citep{2lat,2lac}. 

Out of the 1092 sources designated as AGN,
436 are BL Lacertae objects (BL Lacs), 370 are 
flat-spectrum radio quasar (FSRQs), 12 are radio galaxies, 6 are Seyferts 
and 11 are other AGN.  Despite this important level of
achieved sophistication, the remaining 257 sources 
are designated as active galaxies of uncertain type (AGU) that total
25\% of all AGN. Generally, AGU are positionally coincident with 
flat-spectrum radio sources showing distinctive broad-band blazar characteristics,
but lacking reliable optical measurements \citep{2lac}.  

In order to understand all the intricacies of the AGN population,
it is important to take further steps to assess the nature and redshift of the 
sources classified as AGU. In the past, this has been accomplished 
via a two-step approach. 
The initial classification of an AGN relies on painstakingly 
dedicated optical spectroscopy to help identify unique 
emission or absorption features \citep{shaw}. If no significant 
features are found, the second step consists of 
multi-band photometry to help estimate the redshift of suspected BL Lacs 
\citep{sbarufatti, meisner,rau}.

Without optical spectroscopy, 
one generally does not have sufficient information to determine directly whether an individual source is a BL Lac or a FSRQ.
Unfortunately, optical spectral observations are taxing and can take 
years to complete. 
Ideally, one would like to find a discriminator for distinct
source subclasses that relies solely on readily available observational
characteristics. Recently, 
\citet{massaro} introduced a method that helps recognise gamma-ray 
blazar subclass based on
infrared colours from the Wide-Field Infrared Survey Explorer (WISE).  
Here, we explore the possibility of determining AGN subclass for 
{\it Fermi} sources 
directly from gamma-ray spectral features extracted from the 2FGL. 

In particular, we present results from supervised 
machine learning algorithms, Random
Forests and Support Vector Machines, that are initially 
trained on identified/associated AGN
and subsequently used to infer specific blazar subclass of AGN of uncertain type.
This is a natural
extension of previous machine-learning strategies introduced to predict source
class in unassociated Fermi point sources \citep{class, sibyl, kjlee}.
In Section \ref{algorithm} we describe the machine learning algorithms. The procedure used to
train and test the algorithms is summarised in Section \ref{datasets}, including 
feature selection and creation of datasets. Individual predictions for AGU as well as an extension to
unassociated {\it Fermi} sources 
are presented in Section \ref{results}. Finally, we compare our predictions with \citet{massaro} 
and provide conclusions in Section \ref{discuss}.

\section{Classification algorithms}
\label{algorithm}

The improvement and application of supervised learning algorithms has become a central part 
of the exploration of astrophysical data in a variety of 
contexts, ranging from
object characterisation \citep{ball,class,sibyl}
to variability \citep{richards}. 
In this paper, we employ Support 
Vector Machines (SVMs) and Random Forests (RF) 
that embody two of the most robust supervised learning 
algorithms available today \citep{bloom}. Brief descriptions of both algorithms are given in 
this section. Both RF and SVMs have been extensively 
described in the literature \citep{vapnik, breiman}.

\subsection{Support Vector Machines}
\label{SVMs}
Support Vector Machines (SVMs) have proven to be one of the most  
effective supervised learning algorithms for pattern recognition \citep{vapnik,cortes}.
The underlying rationale behind 
the algorithm seeks to find the optimal margin classifier by constructing
a {\it separating hyperplane} 
that divides the training set and maximises the separation between 
different classes, which can then be used either 
in classification or regression analysis. 

The points lying closer to the boundaries of a certain hyperplane
are called \textit{support vectors}.
The latter determine the minimum distances between the hyperplane and their respective classes,
the so called {\it margin}. The maximisation of the optimal margin is computed 
by taking into account only these vectors, the most representative points 
to construct the classifier. Complex separating surfaces can be introduced
through the use of \textit{kernel} functions, 
which transform the problem into a linear one
in a higher-dimensional space. {\it Polynomial, gaussian} or {\it radial plane}
\textit{kernel} functions are often used. SVMs excel in 
performance handling high-dimensional
data that can also 
incorporate the trade off between training errors and overall
margins parametrized by a scaling factor $\gamma$ and error penalty $C$.

The analysis presented in this work was performed under the 
R programming language. Specifically, we adopted
the \textit{e1071} package as the interface to \textit{libsvm} \citep{libsvm}.
This offers a very fast and efficient SVM application with the option for automatically tuning parameters to
the data and the use of different \textit{kernel} functions.

\subsection{Random Forests}
\label{RFs}

Random Forests is an ensemble classifier that grows a large forest of 
classification trees that independently make class estimation \citep{breiman}.
Each decision tree  selects a number of random input features 
and creates the best split based on a 
{\it out-of-bag} ({\it oob}) random selected set 
of the whole training data sample. Once the decision forest is built, decision thresholds are computed by counting the votes after running the {\it oob} datasets through every tree. 

A RF classifier is ideal for data mining and variable selection as it incorporates 
efficient ways of calculating feature importance in the training set. 
This is achieved by replacing features across classification 
trees with random values and quantifying the effect of the changes. If the result of the 
decisions does not change significantly after these changes, the feature 
has a relatively low importance. On the other hand, if the accuracy rates
change dramatically, a particular feature is deemed as important. 
There 
is no need for cross-validation with a separate testing set as the process 
itself computes accuracy rate internally.

In this work we used the
\textit{randomForest} package \citep{liaw}, which adapts 
the original Random Forests \citep{breiman} for classification 
and regression to the R language. \textit{randomForest} provides 
excellent macros for plotting and tuning. Recently, \citet{sibyl} 
introduced {\it Sibyl} 
a successful classifier that uses Random Forests to predict  
 source class for unassociated {\it Fermi} sources based on 2FGL 
features and it serves as the 
principal formulation for our analysis.

\section{PROCEDURE}
\label{datasets}

Before delving into a formal application of the algorithms, datasets must be 
carefully extracted from the 2FGL and both SVMs and RF 
must be tuned to achieve their best performance. 

\begin{figure*}
\mbox{\subfigure{\includegraphics[width=8.5cm, height=8.5cm]{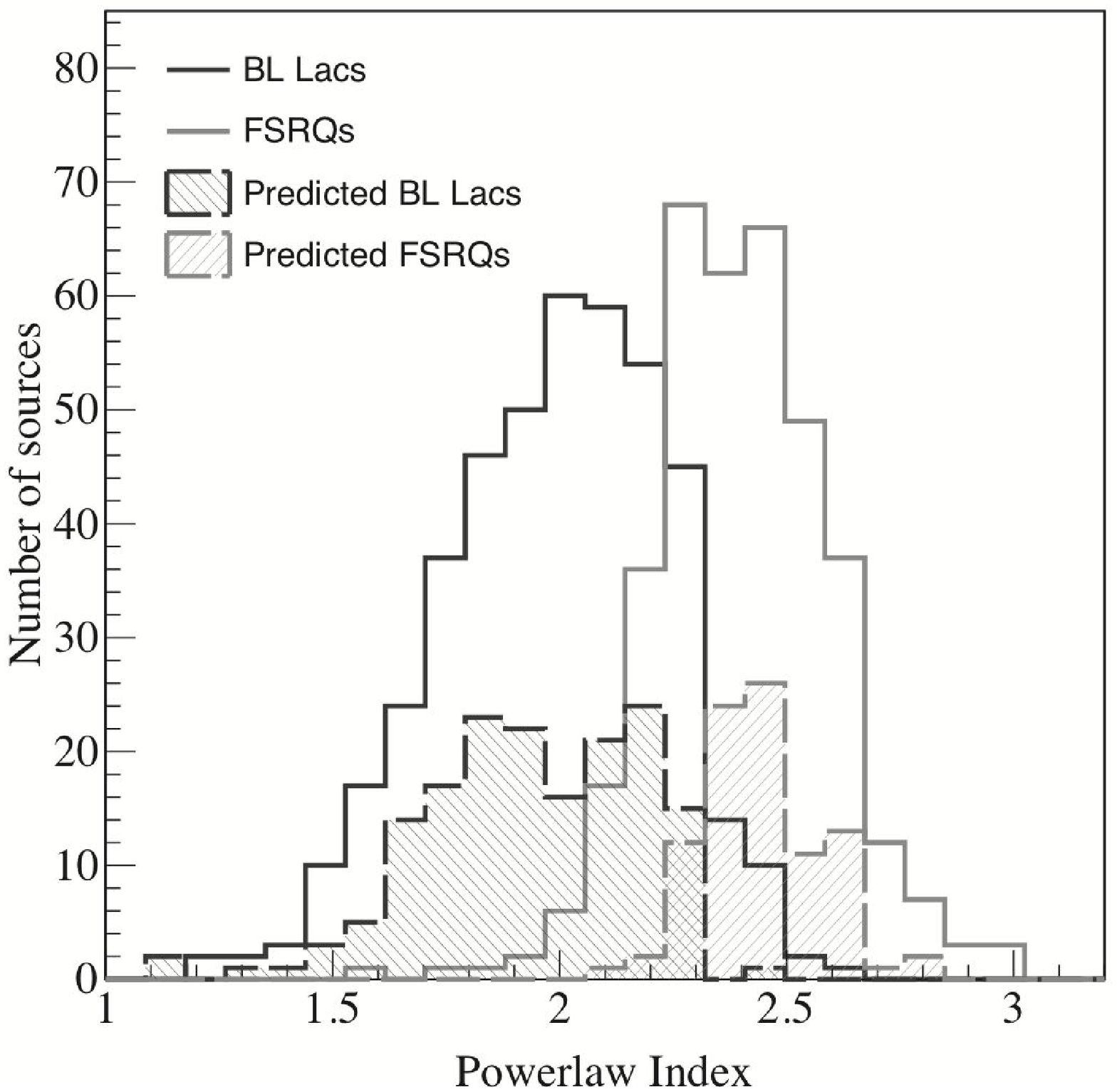}}\quad
\subfigure{\includegraphics[width=8.5cm, height=8.5cm]{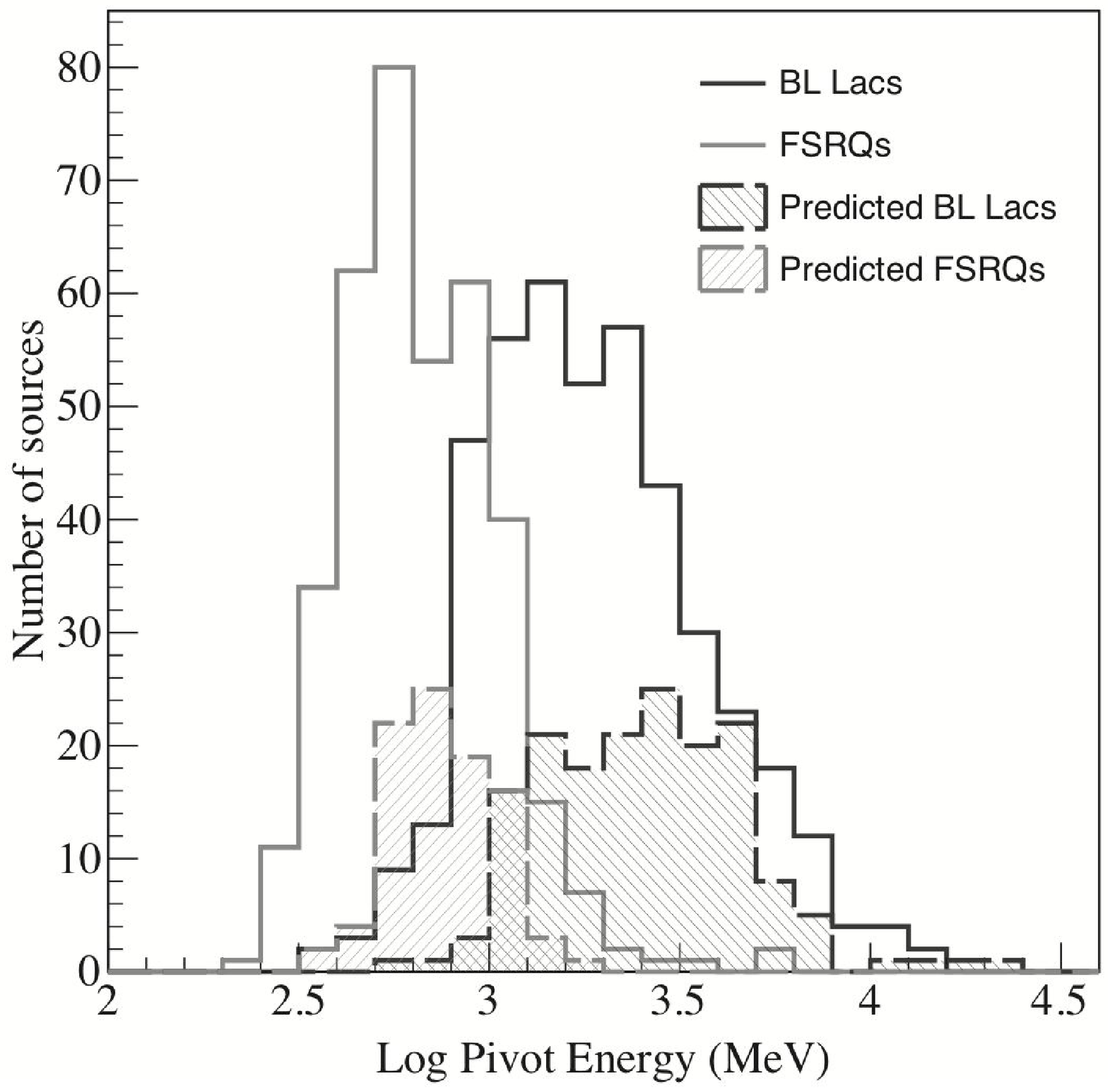}}}
\caption{Distributions of \texttt{Powerlaw Index} (left) and \texttt{Pivot Energy} (right)  
for identified/associated BL Lacs (black) and FSRQs (grey). The filled areas show results for AGU: Predicted BL Lacs (filled dark area) and predicted FSRQs 
(filled light area).}
\label{histograms}
\end{figure*}

\subsection{Construction of the datasets}
\label{datas}

For a proper use of supervised learning algorithms, we need to explore
the feature space in order to find out the variables that 
best capture each class we 
want to determine. In this case, we are interested in building 
classifiers that can distinguish between  
two AGN classes: BL Lacs and FSRQs. 
In the 2FGL,  there are a total set of 1074 
identified/associated AGN objects with the following 
labels: ``bzb'' (BL Lacs), ``bzq'' (FSRQs), ``agn''(other non-blazar AGN) 
and ``agu'' (active galaxies of uncertain type). From this global set, 
we group the identified/associated blazars (``bzb'' and ``bzq'' labels) 
as the  training/testing set of our algorithms.
We end up with a set of 806 sources, divided in a 
fairly balanced manner that includes 370 FSRQs and 436 BL Lacs. 
In addition, we place all the undetermined sources (``agu'' labels)
in a separate dataset consisting
of the 257 objects.
Once the algorithms are trained and tested, 
we will apply the classifiers to the latter. Note that we will initially 
approach our study as a simple binary classification problem that
attempts to distinguish whether an individual AGU is a BL Lac or a FSRQ. It is possible
that other subclasses are represented within the AGU dataset. 
However, additional AGN subclasses
only account for  3\% of the whole AGN sample. Nevertheless,
we will discuss the effects of additional complexity 
later on.

\subsection{Feature selection}
\label{features}

The next step involves choosing from the different gamma-ray spectral features available for each source.
Although the algorithms are not strongly affected by noise, it is relevant to limit misleading
features that might affect the characterisation. Initially, we select
all basic features reported in the 2FGL \citep{2lat}. As in \citet{sibyl},
we supplement these with Hardness Ratios 
($HR_{i}={Flux_{i}-Flux_{j}}/{Flux_{i}+Flux_{j}}$) and Flux Ratios 
($FR_{ij} = Flux_{i}/Flux_{j}$), ending up with a set of 
20 distinct features. Armed with this set of variables, we compute feature 
importance to find those most representative with a robust method
already implemented in the {\it randomForest} package \citep{liaw,sibyl}. 
This process outputs two measures of importance: 
\textit{MeanDecreaseAccuracy} and \textit{MeanDecreaseGini}. 
Both are excellent indicators of feature relevance \citep{breiman}.

Once feature importance measures are computed, we create new sets of data with different number of features by iteratively removing the variables 
with lower \textit{MeanDecreaseAccuracy}, and comparing accuracy rates attained by RF and SVM 
algorithms on these sets. Although RF does not require a tailored  
training/testing analysis to estimate accuracy rates, it is useful 
to compare both algorithms directly with identical training/testing sets.
Through feature selection, we downsize the initial 
20 features to a final set of 9. The final set of variables 
includes (ordered by decreasing \textit{MeanDecreaseGini}) 
\texttt{Powerlaw Index} (76.6), \texttt{Pivot Energy} (59.2), 
\texttt{Flux Density} (27.1), \texttt{Variability Index} (20.1), 
\texttt{Flux1000} (12.6), and four \texttt{Hardness Ratios}: 
$HR_{2}$ (19.4), $HR_{1}$ (17.5), $HR_{3}$ (14.4) \& $HR_{4}$ (10.6). 
Features considered but later discarded include  
\texttt{Spectral Index}, \texttt{Energy Flux}, \texttt{Curvature Index}, 
\texttt{Flux} in five different energy ranges, and 
\texttt{Flux Ratios}.

The top two most representative features for AGN subclass 
determination are \texttt{Powerlaw Index} and \texttt{Pivot Energy}. 
The clean separation between blazars is obvious in
Fig. \ref{histograms} and it intuitively stands 
on observational arguments.
As explained in \citet{2lac}, there is a well established 
spectral difference in the 
LAT energy range between FSRQs and BL Lacs. In general, the AGN inverse Compton (IC) peak is located at lower energies 
for FSRQs and at higher energies for BL Lac objects. Typical values are 1 MeV -- 1
GeV for FSRQ and 100 MeV -- 100 GeV for BL Lacs respectively \citep{seds}. 

The overall effect is that FSRQs show softer
spectra than BL Lacs, and therefore, higher values of 
\texttt{Powerlaw Index}. \texttt{Pivot Energy} is defined as the energy at which the relative
uncertainty on the differential flux is minimal. It is also an estimate of
the point where the covariance of \texttt{Powerlaw Index} and \texttt{Flux Density} is
minimised \citep{2lat}. The relative dominance of lower energy events for FSRQs places the general location of the
\texttt{Pivot energy} at lower energies compared to BL Lac spectra. 
As a result, the difference found in \texttt{Pivot Energy}
between both populations can be understood as the overall effect of the 
spectral characteristics of FSRQs and BL Lacs produced by the
difference on the position of IC peak in the spectral energy 
distribution for both populations.

\subsection{Tuning, training and testing}
\label{training}

Both SVM and RF algorithms require parameter tuning to 
achieve their best performance. In the 
case of SVMs, there is an automatic tuning process {\it best.tune}  
that returns the appropriate  values of $C$ and $\gamma$ for a particular 
\textit{kernel} function and training set.
In order to make a selection, we scanned the classification 
accuracies for different 
\textit{kernel} functions and used the tuned parameters to discriminate 
amongst them. \textit{Linear}, \textit{polynomial}, \textit{sigmoid}, and 
\textit{radial} kernels were tested. For the final training set, we settled on a
\textit{C-classification} \textit{linear} \textit{kernel} with $C = 1$ 
and $\gamma = 0.11$. 
For RF, \textit{tuneRF()} performs an 
automatic search for the most efficient number of features  
used per classification tree for a chosen training set \citep{liaw}. 
Ultimately, we employ 9 spectral features, four variables 
randomly sampled at each split,
and a total of 5000 trees.

After culling our datasets with the chosen features 
and tuning the algorithms for best performance, testing 
is performed to estimate the error of the resulting classification. 
As training set, we use a random selection of $2/3$ of all identified/associated AGN 
and the remaining $1/3$ is used as a testing set. 
To estimate the accuracy rates, we compare the actual source 
class with the class predicted by each classifier. 
For 500 of these training and testing sets, we obtain 
average accuracy rates of $85\%$, adopting a decision threshold of $P > 0.5$ for both RF and SVMs. Note that with such threshold there 
are few ambiguous events since we require both $P_{SVM}$ and $P_{RF}$ must be greater 
than 0.5. If we consider a more conservative condition, for instance 
$P > 0.8$, the accuracy rates improve to $94\%$. In this case, there 
is a bigger fraction of the sample that remains untagged.

For further verification, we also computed rates by 
leaving one object out from the 
training set and using that single object as the testing set. The 
leave-one-out cross validation rate is $85\%$ for common decision 
threshold of $P > 0.5$ and $95\%$ with $P > 0.8$ showing that larger 
training sets do not produce significant increases in accuracy rates.

\section{Results}
\label{results}
\begin{figure}
\hfil
\includegraphics[width=8.5cm, height=8.5cm]{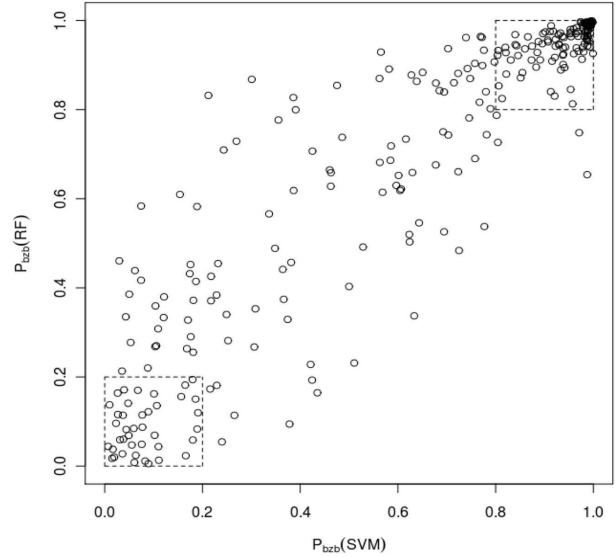}
\hfil
\caption{Decision threshold $P_{bzb}$ obtained with RF versus $P_{bzb}$ estimated by SVM for 257 AGU 
in the 2FGL. Dashed squares contain sources with common decision threshold over 0.8, classified with accuracy rates over 94\%.}
\label{compare}
\end{figure}

Once the classifiers have been trained and tested, 
we apply both algorithms to the set of AGN of uncertain type.
For each of the 257 AGU, the classifiers returns a decision threshold
that an individual object is a BL Lac ($P_{bzb}$) or a FSRQ ($P_{bzq}$), 
where $P_{bzq} = 1 - P_{bzb}$. A fraction of the resulting predictions 
is listed in Table \ref{table1}.
Decision thresholds $P_{bzb}$ calculated with both RF and SVMs are 
shown, as well as a class prediction satisfying   
the condition $P(RF)$  and $P(SVM) > 0.8$. In 
Fig. \ref{compare} we plot $P(RF)$ and $P(SVM)$ obtained with 
each classifier for the 257 sources. Overall, there is an agreement rate of 91\%
between the algorithms. Though there are some discrepancies (for instance RF show higher
BL Lac classification rates than SVMs), the results are outstanding 
considering the distinct underlying assumptions of the algorithms.

\begin{table}
\caption{Predictions for {\it Fermi} AGN of uncertain type in the 2FGL, ordered
 by RA. Threshold values $P_{bzb} <$ 0.2 (in the case of FSRQs) and 
$P_{bzb} >$ 0.8 (in the case of BL Lacs) must be met in both methods.}
\begin{tabular}{l c c l}
\hline
Source             & $P_{bzb}$ (RF) & $P_{bzb}$ (SVM) & Prediction \\
\hline
2FGL    J0001.7-4159    &       0.84    &       0.80    &       bzb     \\
2FGL    J0009.1+5030    &       0.97    &       0.95    &       bzb     \\
2FGL    J0009.9-3206    &       0.53    &       0.57    &       -       \\
2FGL    J0010.5+6556c   &       0.14    &       0.07    &       bzq     \\
2FGL    J0018.8-8154    &       0.69    &       0.80    &       -       \\
2FGL    J0019.4-5645    &       0.16    &       0.04    &       bzq     \\
2FGL    J0022.2-1853    &       0.99    &       1.00    &       bzb     \\
2FGL    J0022.3-5141    &       0.46    &       0.50    &       -       \\
2FGL    J0038.7-2215    &       0.99    &       1.00    &       bzb     \\
2FGL    J0044.7-3702    &       0.06    &       0.04    &       bzq     \\
2FGL    J0045.5+1218    &       0.91    &       0.85    &       bzb     \\
2FGL    J0051.4-6241    &       1.00    &       1.00    &       bzb     \\
2FGL    J0055.0-2454    &       1.00    &       0.99    &       bzb     \\
2FGL    J0056.8-2111    &       0.97    &       0.99    &       bzb     \\
2FGL    J0059.2-0151    &       0.95    &       0.99    &       bzb     \\
2FGL    J0059.7-5700    &       0.03    &       0.02    &       bzq     \\
2FGL    J0103.5+5336    &       0.93    &       0.94    &       bzb     \\
2FGL    J0110.3+6805    &       0.86    &       0.68    &       -       \\
2FGL    J0118.6-4631    &       0.96    &       0.98    &       bzb     \\
2FGL    J0127.2+0324    &       0.98    &       0.99    &       bzb     \\
2FGL    J0131.1+6121    &       0.93    &       0.97    &       bzb     \\
2FGL    J0134.4+2636    &       0.99    &       0.97    &       bzb     \\
2FGL    J0137.7+5811    &       0.44    &       0.38    &       -       \\
2FGL    J0146.6-5206    &       0.95    &       0.92    &       bzb     \\
\hline
\end{tabular}
\\Note: The complete list  of predictions is available at \url{http://www.gae.ucm.es/~thassan/agus.html}.
\label{table1}
\end{table}

\begin{figure*}
\hfil
\includegraphics[width=17.5cm, height=10cm]{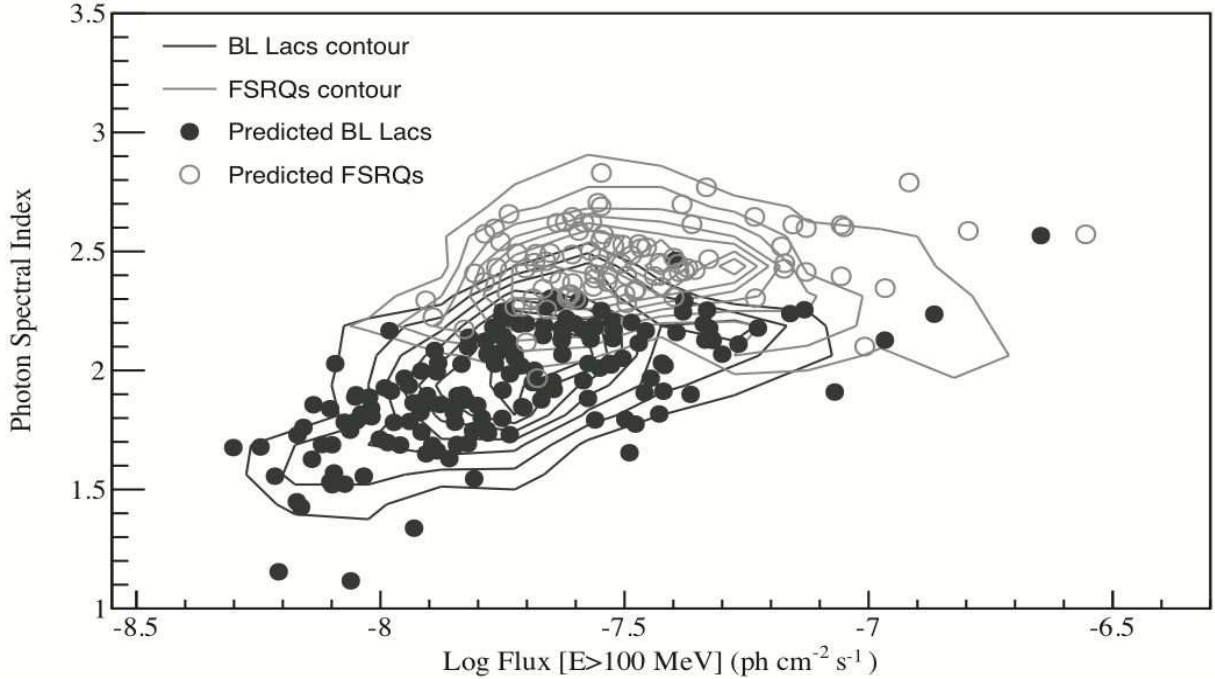}
\hfil
\caption{Photon spectral index versus log flux above 100 MeV for 
identified/associated BL Lacs (dark contour) and FSQRs (light contour). 
Predicted BL Lacs
(filled circles) and predicted FSRQs (open circles) from the AGU dataset
are shown over the contours.}
\label{flux_index}
\end{figure*}

Table \ref{resumen} shows overall numbers sorted according
to different criteria imposed for both RF and SVM. 
In particular we list the predicted number of occurrences in terms of different decision thresholds ($P >$ 0.5, 0.8, and 0.95).
 We include individual algorithms and coincidences,
satisfying said conditions. Combining results from both 
classifiers and requiring $P > 0.5$, 
235 (156 BL Lacs and 79 FSRQs) out of 257 objects are 
consistent with the properties of known gamma-ray blazars. 
In order to place these results in context
with identified/associated {\it Fermi} AGN, Fig. \ref{flux_index} shows 
the photon spectral index versus the
flux ($E >$ 100 MeV) of identified/associated 
BL Lacs and FSRQs overlaid with the 
AGU predictions from this work.

\begin{table}
\caption{Number of predicted AGU sources as a function of decision threshold.}
\centering
\begin{tabular}{||c|c|c|c|c|c|c||} \cline{2-7}
\multicolumn{1}{c}{} &
\multicolumn{2}{|c|}{RF} &
\multicolumn{2}{|c|}{SVMs} &
\multicolumn{2}{|c|}{Both} \\ \cline{2-7}
\multicolumn{1}{c|}{} &
\multicolumn{1}{|c|}{bzb} &
\multicolumn{1}{|c|}{bzq} &
\multicolumn{1}{|c|}{bzb} &
\multicolumn{1}{|c|}{bzq} &
\multicolumn{1}{|c|}{bzb} &
\multicolumn{1}{|c|}{bzq} \\ \hline
$P > 0.5$ & 173 & 84 & 161 & 96 & 156 & 79 \\ \hline
$P > 0.8$ & 129 & 46 & 112 & 63 & 106 & 39 \\ \hline
$P > 0.95$ & 64 & 12 & 64 & 19  & 47 & 5\\ \hline

\end{tabular}
\label{resumen}
\end{table}

\subsection{Outlier detection and potential biases}

Throughout, we have 
assumed that the classification of gamma-ray 
AGN subclass falls along the two main blazar 
categories i.e. BL Lacs and FSRQs. Without final spectroscopy it is 
impossible to rule that other AGN subclasses are present in the AGU sample. 
As commented before, there is a minority of other 
subclasses in the 2FGL including 
Seyferts, radio galaxies and other AGN that have not been considered 
thus far. The main justification for ignoring further atomisation into 
subclasses is 
that blazars account for 97\% of the identified/associated 
AGN sample. However, it is important
to consider that a more complex mixture of AGN subclasses is possible.  
Fortunately, machine-learning algorithms excel at 
separating rare and unique objects from the dataset. 

Adopting the method introduced in \citet{sibyl}, 
we performed a search for AGU outliers that could potentially 
belong to other minority AGN subclasses.
For this purpose, we computed the 
outlying measure of each object 
defined as the reciprocal sum
of the squared proximities to all objects within its class. 
Outliers are defined as objects having small proximities to the
rest of objects. Practically, 
RF returns proximities $prox(n,k)$ that represent 
the fraction of trees in which elements $n$ and ${k}$
fall in the same terminal node \citep{breiman,liaw}.
Generally, anomalies are identified with outlier measures larger than 10. 
No source was found with such values, as 
a result we conclude that there is
no clear evidence of outliers in the AGU dataset. 
For completeness, we note that the highest values in the 
dataset correspond to 2FGL J1825.1--5231,
2FGL J1816.7--4942, and 2FGL J0022.3--5141 respectively.

We constrained this possibility further by 
retraining and testing the SVMs and RF algorithms 
with the full range of associated AGN subclasses present in the 2FGL. Given the
size of the minority subclasses, care was taken to
weight the classes appropriately to compensate the differences 
in the training sets \citep{sibyl}. Taking into account additional AGN 
subclasses, we find that at most 
11 objects might belong 
to other AGN subclasses ($P <$ 0.6). Therefore, there is no strong 
indication of contamination from additional subclasses. 
Taken together, both approaches limit the presence of   
other AGN subclasses in the AGU dataset. It is possible that the result
simply reflects the 
small number statistics of additional AGN subclasses. A full
characterization might improve in the future as {\it Fermi} expands its
source catalogue.

\subsection{Application to unassociated {\it Fermi} objects}

In \citet{sibyl}, we introduced class predictions for the sample of
unassociated {\it Fermi} sources at $|b| \geq 10^\circ$. In that
initial approach, we simply considered sorting sources in broad
AGN and pulsar categories. Given our success with further AGN subclasses,  
it is interesting to extend the approach to all unassociated
{\it Fermi} sources tagged as AGN.
Using the same optimised models, we
apply the algorithms to the 216 sources predicted as AGN in \citet{sibyl}.
The resulting predictions are shown in Table 
\ref{table_sibyl} with the same  
conditions adopted earlier.
In this case, only 30\% of the sources reach 
decision thresholds larger than $P >$ 0.8
in both RF and SVM.

\begin{table}
\caption{Predictions for  unassociated \textit{Fermi} objects tagged as AGN by \citet{sibyl}, ordered by RA.}
\begin{tabular}{l c c l}
\hline
Source             & $P_{bzb}$ (RF) & $P_{bzb}$ (SVM) & Prediction \\
\hline
2FGL	J0004.2+2208	&	0.15	&	0.11	&	bzq	\\
2FGL	J0014.3-0509	&	0.37	&	0.19	&	-	\\
2FGL	J0031.0+0724	&	0.97	&	0.94	&	bzb	\\
2FGL	J0032.7-5521	&	0.41	&	0.28	&	-	\\
2FGL	J0039.1+4331	&	0.87	&	0.99	&	bzb	\\
2FGL	J0048.8-6347	&	0.91	&	0.76	&	-	\\
2FGL	J0102.2+0943	&	0.90	&	0.89	&	bzb	\\
2FGL	J0103.8+1324	&	0.94	&	0.95	&	bzb	\\
2FGL	J0116.6-6153	&	0.97	&	0.99	&	bzb	\\
2FGL	J0124.6-2322	&	0.49	&	0.66	&	-	\\
2FGL	J0129.4+2618	&	0.19	&	0.05	&	bzq	\\
2FGL	J0133.4-4408	&	0.63	&	0.73	&	-	\\
2FGL	J0143.6-5844	&	1.00	&	0.99	&	bzb	\\
2FGL	J0158.4+0107	&	0.36	&	0.26	&	-	\\
2FGL	J0158.6+8558	&	0.06	&	0.07	&	bzq	\\
2FGL	J0200.4-4105	&	0.98	&	0.99	&	bzb	\\
2FGL	J0221.2+2516	&	0.99	&	0.99	&	bzb	\\
2FGL	J0226.1+0943	&	0.66	&	0.76	&	-	\\
2FGL	J0227.7+2249	&	0.89	&	0.95	&	bzb	\\
2FGL	J0239.5+1324	&	0.99	&	0.95	&	bzb	\\
2FGL	J0251.0+2557	&	0.37	&	0.19	&	-	\\
2FGL	J0305.0-1602	&	0.99	&	1.00	&	bzb	\\
2FGL	J0312.5-0914	&	0.93	&	0.69	&	-	\\
2FGL	J0312.8+2013	&	0.91	&	0.97	&	bzb	\\

\hline
\end{tabular}
\\Note: The complete list  of predictions is available at \url{http://www.gae.ucm.es/~thassan/agus.html}.
\label{table_sibyl}
\end{table}

\section{Discussion and Conclusions}
\label{discuss}
We have used RF and SVM classifiers 
to predict specific source subclasses for gamma-ray 
AGN of uncertain type, by learning from features extracted from 
associated AGN in the 2FGL. Both algorithms are extremely successful in 
capturing the properties of gamma-ray AGN reaching
accuracy rates of  85\%. This effort allows us to 
show that 235 out of 269 AGN of uncertain type have properties
consistent with gamma-ray BL Lacs and FSRQs, with decision thresholds over 0.8.
Comparison of these predictions with the sample of associated AGN 
verify that we are indeed tracing similar populations (Fig. \ref{flux_index}).
Nevertheless, without high-quality spectral observations, 
final counterpart 
association will have to wait for dedicated optical spectroscopy.

Apart from internal training and testing, we can 
cross-match our results with a  
recent study showing that blazars can be recognised and separated from 
other extragalactic sources using infrared colours \citep{massaro}. 
Class characterisation has been done for {\it Fermi} 
AGN of uncertain type taking advantage of this 
\textit{total strip parameter} traced by BL Lacs and FSRQs. 
The possibility of comparing our results with the 
source classes inferred from IR colours is ideal, as both methods 
are independent. 
For a subset of 54 overlapping sources listed in \citet{massaro},
our predictions match in 85\% of the objects with  
the $P > 0.5$ decision threshold, and 
the agreement rate improves to 93\% for the 33 objects satisfying the $P > 0.8$ condition. 
The excellent agreement suggests that our method is viable and that infrared colours can not only 
recognise generic blazars but also provide information about specific
blazar subclass i.e. BL Lac or FSRQ. More importantly, 
this cross-validation reinforces the power and possibilities of 
machine-learning algorithms as source classifiers in 
gamma-ray astrophysics. 

Even though the initial approach aimed to distinguish 
between BL Lacs and FSRQs, we have also considered 
the possibility that other subclasses are represented within the AGU dataset. 
No clear outliers have been found within the latter. 
Training and testing after taking into consideration additional subclasses
finds only 11 objects ($P < 0.6$) that might have been missed with a 
binary classification. This is consistent with findings indicating 
that additional AGN subclasses (Seyferts, radio galaxies and other AGN) 
account for a 3\% of the whole AGN sample. There might be a small bias 
introduced by the relative rarity of minority objects. Nevertheless, 
AGN of unknown type are 
most likely dominated by BL Lac or FSRQ, in agreement with \citet{massaro}.

The clear intent of this effort 
is to characterise the entire gamma-ray population. 
We expect that these results can help 
observers in future spectral and photometric
endeavours aimed at classifying the entire AGN counterpart sample. 
Additionally, our work can help  
discriminate targets 
for follow-up studies of AGN  
at even higher gamma-ray energies with 
ground-based imaging air Cherenkov telescopes 
(MAGIC, H.E.S.S., VERITAS). Viewing forward, gamma-ray spectral 
features will be nicely complemented with the future 
Cherenkov Telescope Array (CTA), expected
to increase spectral coverage and sensitivity \citep{cta}.  
The design of future survey pointing strategies 
for CTA \citep{dubus} will also benefit from object 
lists such as the one presented 
in this work by boosting the AGN target pool available. 
In the shorter term, an obvious improvement that lies ahead 
is to incorporate multi-wavelength (radio, optical, X-ray) 
entries to complement individual
classifying features. This is an area that we are currently 
investigating.

\section*{Acknowledgments}

The authors acknowledge the support of the Spanish MINECO under project FPA2010-22056-C06-06 and the German Ministry for Education and Research (BMBF). N.M. acknowledges support from the Spanish government.  
through a Ram\'on y Cajal fellowship.
We also thank the referee for useful suggestions and 
comments on the manuscript.

\label{lastpage}
\end{document}